\documentclass[nofootinbib,superscriptaddress,
showpacs]{revtex4}
\usepackage{graphicx}
\begin{document}
\title{Atom laser dynamics in a tight-waveguide}

\author{A. del Campo}
\email{adolfo.delcampo@ehu.es}
\address{Departamento de Qu\'imica-F\'isica,
UPV-EHU, Apartado. 644, Bilbao, Spain}
\author{I. Lizuain}
\address{Departamento de Qu\'imica-F\'isica,
UPV-EHU, Apartado. 644, Bilbao, Spain}
\author{M. Pons}
\address{Departamento de F\'isica Aplicada I,
E.U.I.T. de Minas y Obras P\'ublicas, UPV-EHU, 48901 Barakaldo, Spain}
\author{J. G. Muga}
\address{Departamento de Qu\'imica-F\'isica,
UPV-EHU, Apartado. 644, Bilbao, Spain}
\author{M. Moshinsky}
\address{$^3$ Instituto de F\'isica, Universidad Nacional Aut\'onoma de M\'exico, 
 Apartado Postal 20-364, 01000 M\'exico D.F., M\'exico}

\def\la{\langle} 
\def\ra{\rangle}
\def\om{\omega}
\def\Om{\Omega}
\def\vep{\varepsilon}
\def\wh{\widehat}
\def\tr{\rm{Tr}}
\def\da{\dagger}
\newcommand{\beq}{\begin{equation}}
\newcommand{\eeq}{\end{equation}}
\newcommand{\beqa}{\begin{eqnarray}}
\newcommand{\eeqa}{\end{eqnarray}}
\newcommand{\bV}{{\bf V}}
\newcommand{\bK}{{\bf K}}
\newcommand{\bG}{{\bf G}}
\begin{abstract}

We study the transient dynamics that arise during the formation of an atom laser beam in a tight waveguide.
The time dependent density profile develops a series of wiggles which are related 
to the diffraction in time phenomenon.
The apodization of matter waves, which relies on the use of smooth aperture functions, allows to 
suppress such oscillations in a time interval, after which there is a revival of the diffraction in time.  
The revival time scale is directly related to the inverse of the harmonic trap frequency for the atom reservoir.

\end{abstract}
\maketitle
Atom lasers have recently been created in the laboratory and stand up for 
their applications in atom lithography and interferometry. Similar to their light analogues, they consist in 
 coherent and directed (matter-) wave beams, which are extracted from a Bose-Einstein condensate (BEC). 

Different schemes have been proposed to out-couple the atoms from the BEC reservoir.   
One of the prototypes uses two-photon Raman excitation   
to create ``output coupled'' atomic pulses with well defined mean momentum  
which overlap and form a quasi-continuous beam \cite{Phillips99,Phillips00} (For other 
approaches see e.g. \cite{Ketterle97} or \cite{CMRVPG06}).
The features of the beam profile are of outmost importance for applications 
\cite{KBMHE05,RGLCFJBA06,KP06}, and 
depend on the pulse shape, duration, emission frequency, and initial phase \cite{DMM07}.  

We study the dynamics of a beam created by multiple overlapping pulses, 
transversely confined in a tight waveguide and with
negligible interactions between the out-coupled atoms.
It is found that an oscillatory pattern appears in the density profile of the beam which is closely 
related to the diffraction in time phenomenon. If the switching process is slow, 
such oscillations are suppressed on a time window after which there is a revival 
of the diffraction in time.  The time scale in which the spatial fringes appear 
is related to the the spatial width of the out-coupled pulses.

\section{Turning-on the source\label{turning-on}}

Following  the recent realization of an atom laser in a waveguide \cite{Guerin06}, 
we shall consider the dynamics of matter waves with momentum $\hbar k_0$ 
released in a waveguide of frequency $\om_{\perp}$, 
tight enough so that the transverse excitation quanta $\hbar\om_{\perp}$ 
is much larger than any other relevant energy scales. 
Under such assumption, the radial degrees of freedom, being frozen, play no role and the system 
becomes effectively one dimensional.
Moreover if the setup is horizontal, the gravitational field is constant, 
avoiding the decrement of the de Broglie wavelength $\lambda_{dB}=2\pi/k_0$ 
due to the downward acceleration observed in vertical setups \cite{Guerin06}. 
However, the effects of an arbitrary time-dependent linear potential, 
of interest to achieve controlled transport, can be included following \cite{DM06a}.

The time evolution along the longitudinal axis can be tackled using 
``source'' boundary conditions of the form
\beq\label{source0} 
\psi(x=0,t)=\chi(t)\,e^{-i\om_0 t}, \qquad \forall t>0,
\eeq
where the ``carrier'' frequency is related to the characteristic wavenumber of the beam by $\om_0=\hbar k_0^2/2m$, 
and the aperture function $\chi(t)$ modulates the wave amplitude 
at the source. Such localized sources \cite{BT98,MB00} have been used for understanding tunneling 
dynamics and related time quantities such as the tunneling times \cite{MSE02}, transient 
effects in neutron optics \cite{GG84,FGG88,HFGGGW98}, and diffraction of atoms both in 
time and space domains \cite{BZ97,DM05}. 
The connection and equivalence between ``source'' boundary conditions, 
in which the state is specified at a single point at all times and the usual initial value problems, 
in which the state is specified  for all points at a single time, was studied in \cite{BEM01}. 

Whenever the modulation of the source is sudden 
($\chi(t)=\Theta(t)$, where $\Theta(t)$ is the Heaviside step function) 
the resulting dynamics exhibits a wiggly density profile that, by analogy with the 
diffraction of light from a semi-infinite plane, became known as 
diffraction in time \cite{Moshinsky52,Moshinsky76}.
%
\begin{figure} 
\begin{center}
\includegraphics[height=6cm,angle=0]{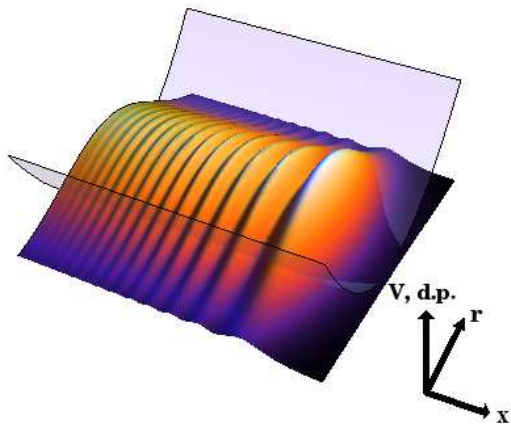}
\caption{\label{sit}
Schematic snapshot of a suddenly turned-on atom beam propagating rightwards along a waveguide and exhibiting diffraction in time.
The wavefront of the density profile (d.p.) develops a set of oscillations, which is the hallmark feature of  diffraction in time, 
in contrast with the uniform classical case.
Whenever the harmonic transverse confinement V is tight enough (represented by the concave sheet), the radial degrees of freedom are frozen out and the dynamics becomes essentially one dimensional, along the longitudinal axis $x$.
}
\end{center}
\end{figure}
%
Figure \ref{sit} shows a snapshot of the density profile of a suddenly turned-on atom beam,  with the oscillatory pattern 
which is the tale-telling sign of diffraction in time. 
This singular matter wave phenomenon has been observed in a wide variety of systems including ultracold neutrons 
\cite{HFGGGW98} and atoms \cite{atomdit}, electrons \cite{Lindner05}, and even Bose-Einstein 
condensates \cite{CMPL05}. Such experimental flurry together with the simplicity of Moshinsky's result has spurred a 
series of works dealing with more complex setups \cite{Kleber94,MB00,GM05,DM05,DM06a}. 

Here, instead of sudden switching we shall consider smooth aperture functions, 
i.e., ``apodization'', a technique well-known in Fourier optics to avoid the diffraction effect
at the expense of broadening the energy distribution \cite{Fowles68}.
Manifold applications of this technique have also been found in time filters for signal analysis \cite{BT59}.  
\section{Smooth switching: apodization and diffraction in time}\label{switching}

In this section we shall study the modification of matter-wave dynamics for slow aperture functions 
with a finite switching time $\tau$. We introduce the continuous switching function
\begin{displaymath}
\label{switch}
\chi^{s}(t)=
\cases{
 0, \qquad t<0\\
\chi(t), \quad 0\leq t<\tau\\
1, \qquad t\geq\tau},
\end{displaymath}
where the superscript $s$ denotes that the source remain open for all $t>0$ (a {\it s}witch).
The family for which $\chi(t)=\sin^n\Om_s t$ with $n=0,1,2$ and $\Om_s=\pi/2\tau$ was considered in \cite{DMM07}, 
where it was shown that for fixed $\tau$, the apodization of the beam increases with the smoothness of $\chi(t)$, 
this is, with $n$. Note that for $n=0$, one recovers the sudden aperture $\chi_0(t)=\Theta(t)$ which 
maximizes the diffraction-in-time fringes. We shall consider a sine-square modulation in what follows ($n=2$) 
and concentrate on the dependence on $\tau$.

%
\begin{figure} 
\begin{center}
\includegraphics[width=7.5cm,angle=0]{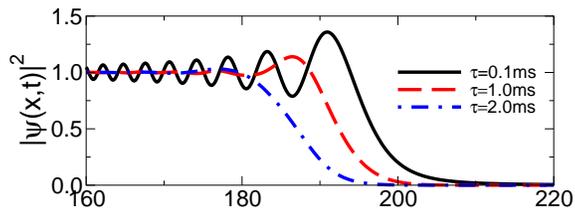}
\caption{\label{apodization}
Apodization in time. Removal of the sudden approximation in the switch of a matter wave 
source leads to the suppression of oscillations in the beam profile characteristic of the diffraction in time. 
The $^{87}$Rb beam profile is shown at $t=20$ms, and $\hbar k_0/m=1$cm/s. 
For increasing $\tau$ the amplitude of the oscillations diminishes and the signal is delayed.
}
\end{center}
\end{figure}
%

Increasing the switching time $\tau$  leads to an apodization of the oscillatory pattern. As a result 
the fringes of the beam profile are washed out, see Fig. \ref{apodization}. 
In this sense, the effect is tantamount to that of a finite band-width source \cite{MB00}, 
and indeed this is the key to apodization, the suppression of diffraction is achieved broadening 
 the energy distribution \cite{BT59,Fowles68}.
It is also analogous to the suppression of the current transients in 
tunneling through a time-dependent barrier. Whenever the modulation of the potential is slow enough, 
the resulting dynamics becomes adiabatic and the tunneling current tends 
to the quasistatic equilibrium current \cite{SK88}.  

\begin{figure} 
\begin{center}
\includegraphics[width=7.5cm,angle=0]{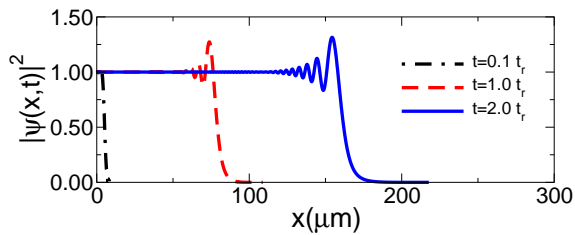}
\caption{\label{revival}
Revival of the diffraction in time.
During the time evolution of an apodized source, there is a revival of the diffraction 
in time in the time scale $t_r=\om_0\tau^2$ ($\tau=1$ms, $\hbar k_0/m=0.5$cm/s, 
and $t_r=16.8$ms for a $^{87}$Rb source which is switched on following a sine-square function).
}
\end{center}
\end{figure}

\begin{figure} 
\begin{center}
\includegraphics[height=6cm,angle=0]{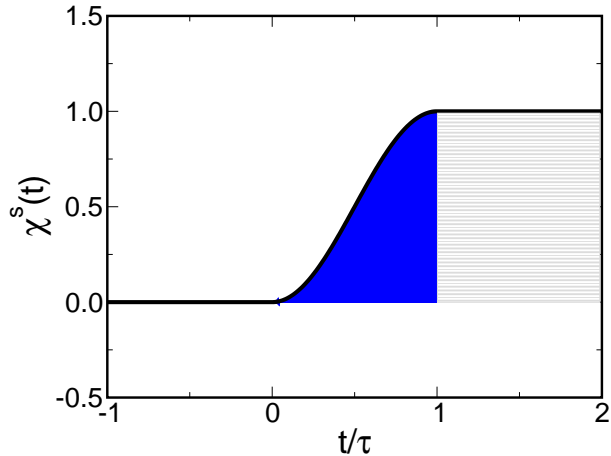}
\caption{\label{cap}
Switching aperture function decomposition. In a switching process the aperture 
function has a finite transient component 
responsible for the apodization (filled area). 
Such apodization cap is followed by a step excitation for $t\geq\tau$  (shaded area)
which leads eventually to the revival of diffraction in time. 
}
\end{center}
\end{figure}

However, the apodization lasts only for a finite time 
after which a {\it revival} of the diffraction in time occurs, as shown in Fig. \ref{revival}. 
The intuitive explanation is that the intensity of the signal from the apodization ``cap'', 
see Fig. \ref{cap}, decays with time whereas the intensity of the main 
signal (coming from the step excitation) remains constant. 
For sufficiently large times, the main signal, carrying its 
diffraction-in-time phenomenon, overwhelms the effect of the 
small cap. 
Indeed, from the linearity of the Schr\"odinger equation, 
it follows that the wavefunction is the sum of a ``half-pulse'' term 
associated with the switch released during the interval $[0,\tau]$ and a semi-infinite beam 
suddenly turned on at time $t=\tau$.
%
%

One can estimate the revival time by considering the 
momentum width induced by the apodization 
cap, $\Delta p\approx
\sqrt{2m\hbar(\om_0+\Om_s)}-\sqrt{2m\hbar(\om_0-\Om_s)}
\approx\sqrt{2m\hbar\om_0}\frac{\pi}{2\om_0\tau}$, 
and ballistic spread of the cap half-pulse 
(we have checked numerically 
that the classical dispersion relation for such pulse $\Delta x(t)\simeq\Delta x(0)+\Delta pt_r/m$ holds). 
%
%

The ratio of the spatial widths corresponding to the initial 
and evolved cap $\psi_{\chi}^{(1)}$ is given by
\beq
R=\frac{\Delta x(0)}{\Delta x(t_r)}\approx\frac{\Delta x(0)}{\Delta x(0)+\Delta pt_r/m}.
\eeq
Taking $\Delta x(0)=\sqrt{2\hbar\om_0/m}\tau$ and imposing $R=1/2$ leads to 
the expression for a revival time scale 
\beq
t_r\approx \om_0\tau^2,
\eeq
which is analogous to the Rayleigh distance of classical diffraction theory 
but in the time domain \cite{Brooker03}.
The smoothing effect of the apodizing cap 
cannot hold for times much longer than $t_r$.
This is shown in Fig. \ref{revival}, where the initially 
apodized beam profile develops eventually spatial fringes, which reach the maximum value associated 
with the sudden switching at $t\approx 2t_r$.

\section{Multiple pulses}\label{mp}

The matter-wave dynamics of a single pulse obtained from a source by means of 
an aperture function $\chi^{(1)}(t)$ has been described in \cite{DM05,DMM07} 
(hereafter, the superscript denotes the number of pulses in the aperture function). 
Both the aperture function $\chi^{(1)}(t)$, and time-dependent wavefunction of 
the associated pulse $\psi^{(1)}(x,t)$, allow a simple description of 
matter wave sources periodically modulated in time.

More precisely, the $N$-pulse aperture function $\chi^{(N)}$ can be written as a convolution of 
$\chi^{(1)}(t)$ with a grating function $g(t)$.
If the grating function $g(t)$ is chosen to be a finite Dirac comb 
 $g(t)=\sum_{j=0}^{N-1}\delta(t-j T)$, it follows that 
\beq
\chi^{(N)}(t)=\chi^{(1)}(t)*g(t)=\int dt'\chi^{(1)}(t-t')g(t')=\sum_{j=0}^{N-1}\chi^{(1)}(t-jT).
\eeq
Such modulation describes the formation of $N$ consecutive $\chi^{(1)}$-pulses, 
each of them of duration $\tau$ and with an ``emission rate'' (number of pulses per unit time) $1/T$.

This mathematical framework can be employed to describe an atom laser
in the non-interaction limit. The experimental prescription for an atom laser 
depends on the out-coupling mechanism.
Here we shall focus on the scheme described in \cite{Phillips99,Phillips00}.
The pulses are obtained from a condensate trapped in a MOT, and a well-defined momentum is 
imparted on each of them at the instant of their creation through a
stimulated Raman process. The result is that each of the pulses  $\psi^{(1)}$ 
does not have a memory phase, the wavefunction describing the coherent atom laser being then  
\beq
\phi^{(N)}(x,k_0,t)=\sum_{j=0}^{N-1}\psi^{(1)}(x,k_0,t-jT;\tau)
\Theta(t-jT).
\eeq
The Heaviside function $\Theta(t-jT)$ implies that the $j$-th 
 pulse starts to emerge only after $jT$. It is interesting to impose a 
relation between the out-coupling period $T$ and 
the kinetic energy imparted to each pulse $\hbar\om_0$, such that $\om_0 T=l\pi$, 
where if $l$ is chosen an even (odd) integer 
the interference in the beam profile $|\phi^{(N)}|^2$ 
is constructive (destructive).

The diffraction in time is a coherent quantum dynamical effect, 
we may expect it to be affected by noise in the environment. 
If the phases between different pulses are allowed to fluctuate, the 
interference pattern is affected \cite{VVDHR06}.
Moreover, if there is no phase coherence between different pulses, 
the resulting density profile is the incoherent sum 
of the single-pulse densities, 
$\rho(x,t)=\sum_{j}\vert\psi^{(1)}(x,k_0,t-jT;\tau)\vert^{2}\Theta(t-jT)$, 
which we shall use as a reference case.

\begin{figure} 
\begin{center}
\includegraphics[height=6cm,angle=0]{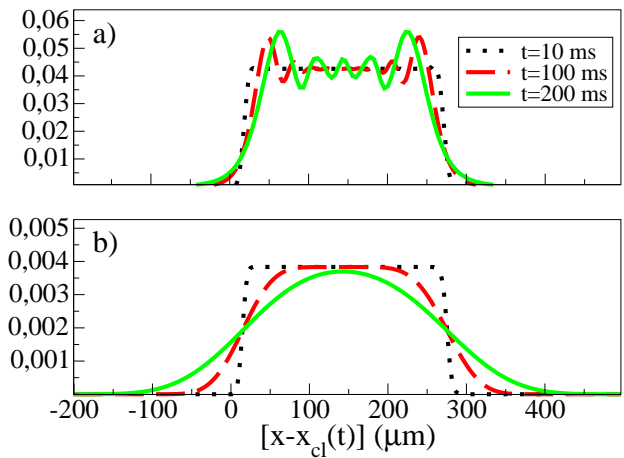}
\hspace*{.2cm}
\caption{
\label{atltime}
Time evolution of the density profile of a $^{23}$Na atom laser beam modulated by a sine-square function 
with $\tau=0.833$ms, $T=0.05$ms, moving at $3.73$cm/s in the a) coherently constructive, b) incoherent case.
Note that $x_{cl}=\hbar k_0 t/m$ is the classical trajectory. In all cases the norm is relative to the 
incoherent case, as explained in the text. The revival time is $174.4$ ms.
}
\end{center}
\end{figure}

One advantage of employing stimulated Raman pulses is that any desired fraction of atoms 
can be extracted from the condensate. Let us define $s$ as the number of atoms out-coupled 
in a single pulse, $\psi^{(1)}$. The total norm for the $N$-pulse incoherent atom laser is simply $\mathcal{N}_{inc}=sN$.
A remarkable fact for coherent sources is that the total number of atoms $\mathcal{N}_c$
out-coupled from the Bose-Einstein condensate reservoir depends on the nature of the interference \cite{Felipe}.
Therefore, in Fig.  \ref{atltime} we shall consider the signal relative to the incoherent case, 
namely, $|\phi^{N}(x,t)|^2/\mathcal{N}_{inc}$.
%
 
%
Note the difference in the time evolution between coherently 
constructive and incoherent beams in Fig. \ref{atltime}.
At short times both present a desirable saturation in the probability density of the beam.
However, in the former case a revival of the diffraction in time phenomena takes place in a 
time scale $\om_0\tau^{2}$, similar to the smooth switching considered in section \ref{switching}, 
whereas the later develops a bell-shape profile.

\section{Revival of the diffraction in time in an atom laser}


In an atom laser, the revival time can be related to the frequency of the BEC reservoir trap $\om_{trap}$, 
which determines the width of the out-coupled pulses.

First we note that whenever the overlap amongst different pulses 
is coherently constructive or with phase memory (as in the periodic chopping of a beam), 
an effective single-pulse aperture function can be introduced. 
Indeed, if $\tau/T\gg1$ as in the actual experiments \cite{Phillips99,Phillips00}, 
it follows that an effective single-pulse aperture function of duration $\tau_{eff}=(N-1)T+\tau$ can be introduced, 
%
 with a plateau in the time interval $[\tau,NT-\tau]$ and both a front and a back cap of duration $\tau$ associated with the switching on and off. 
These caps determine the revival time $\om_0\tau^{2}$. 

Assume that atoms are out-coupled from the $n$-th longitudinal mode of the reservoir trap, whose 
spatial width is $\xi_n=[(n+1/2)\hbar/m\om_{trap}]^{1/2}$. Using $\om_0=\hbar k_{0}^2/2m$, 
and the semiclassical relation $\tau_n=m\xi_n/\hbar k_0$,
the revival time becomes 
\beqa
t_r^{(n)}=\left(n+\frac{1}{2}\right)\frac{1}{2\om_{trap}}.
\eeqa
A non-interacting Bose-Einstein condensate, one can take $n=0$. 
This is a remarkable relation which states that diffraction in time manifest 
itself since the beginning of the experiment in the limit of very tight traps which lead to narrow pulses.
However, if the Bose-Einstein condensate trap is broad, 
so will be the out-coupled pulses and apodization will suppress the oscillations 
on the density profile for times smaller than the revival time.
\section{Discussion}

We have studied the dynamics of non-interacting boson sources which are 
modulated in time with an arbitrary aperture function. 
Whenever the modulation is sudden, the spreading of the source exhibits an
oscillatory pattern in the density profile, which is the tale-telling sign 
of the quantum diffraction in time phenomenon. 
For smooth aperture functions the oscillations 
are suppressed, thus bringing to the quantum domain the classical apodization techniques 
usual in signal analysis and Fourier optics.
However, for long or multiple overlapping pulses 
the smoothing is limited by a time scale $t_r$, associated with a revival 
of the diffraction-in-time phenomenon.
Multiple pulse formation has been examined for different possible cases. 
For constructive or for random averaged phases, a saturation effect occurs 
providing a flat intensity profile before $t_r$, which later  develops 
spatial fringes and a bell-shape profile respectively. 

Though our results consider the dynamics under strong transverse confinement, 
they can be extended beyond one dimension. Indeed, the effect of the dimensionality 
on the diffraction in time phenomenon, have already been considered in a different context, 
namely, the expansion of a single particle from cylindrical and spherical traps 
\cite{Godoy03,Godoy05}. The result is that the amplitude of the oscillations decreases 
in two and three dimensional setups in comparison with the dynamics along a tight-waveguide presented here.

Finally, we have restricted this work to the case of non-interacting bosons.
However, the presence of interactions between the out-coupled atoms 
is expected to reduce the amplitude of the diffraction in time phenomenon, 
as suggested by previous studies of quantum transients in the strongly interacting regime \cite{DM06}.

\begin{acknowledgments}
A. C. acknowledges the hospitality of the members of IF-UNAM during the completion of this work and financial support by the Basque Government (BFI04.479). The work has been supported by CONACYT (40527F), Ministerio de Educaci\'on y Ciencia (BFM2003-01003, FIS2005-01369 and FIS2006-10268-C03-01), and UPV-EHU (00039.310-15968/2004).
\end{acknowledgments}

\end{document}